\documentclass[aip, amsmath, amssymb, reprint]{revtex4-1}
\usepackage{graphicx}
\usepackage{dcolumn}
\usepackage{bm}
\usepackage[utf8]{inputenc}
\usepackage[T1]{fontenc}
\usepackage{mathptmx}
\usepackage{etoolbox}
\usepackage{color,soul}
\usepackage{xcolor}
\usepackage{lipsum}
\usepackage{array}
\usepackage{multirow}
\usepackage{comment}
\usepackage{hhline}
\usepackage{caption}

\renewcommand{\arraystretch}{1.15}
\draft

\begin{document}

\title{Automation in quantum logic experiments with cold molecular ions}

\author{Richard Karl}
\author{Meissa Diouf}
\author{Aleksandr Shlykov}
\author{Mikolaj Roguski}
\author{Stefan Willitsch}
\email[]{stefan.willitsch@unibas.ch}

\affiliation{Department of Chemistry, University of Basel, Klingelbergstrasse 80, Basel 4056, Switzerland}

\begin{abstract}
Modern experiments with cold molecular ions have reached a high degree of complexity requiring frequent sample preparation, state initialization and protocol execution while demanding precise control over multiple devices and laser sources. To maintain a high experimental duty cycle and robust measurement conditions, automation becomes essential. We present a fully automated control system for the preparation of trapped state-selected molecular ions and subsequent quantum logic-based experiments. Adaptive feedback routines based on real-time image analysis introduce and identify single molecular ions in atomic-ion Coulomb crystals. By appropriate manipulation of the trapping potentials, excess atomic ions are released from the trap to produce dual-species two-ion strings, here Ca$^+-$N$_2^+$. After mass and state identification of the molecular ion, nanosecond-level synchronization of laser pulses employing the Sinara/ARTIQ framework and real-time data analysis enable quantum-logic-spectroscopic measurements. The present automated control system enables robust, unsupervised operation over extended periods resulting in an increase of the number of experimentation cycles by about a factor of ten compared to manual operation and a factor of about eight in loaded molecules in typical practical situations. The modular, distributed design of the system provides a scalable blueprint for similar molecular-ion experiments.
\end{abstract}


\maketitle

\section{Introduction}
Modern atomic, molecular, and optical physics experiments involve a wide range of instruments that need to operate in precisely controlled sequences. Integrated control systems capable of interfacing with multiple devices\cite{Yu19a} and executing synchronized tasks on timescales ranging from hours to nanoseconds\cite{Perego18a} are crucial for state-of-the-art measurements. Moreover, experimental setups tend to become increasingly intricate and difficult to manage\cite{Volponi24a}, especially when they operate for many years with continuous upgrades by multiple users.\\
Some of these challenges can be addressed by following a set of design strategies when building control systems for such experiments. A modular design with automated subroutines allows for easier integration of new features and flexibility to pursue various research objectives\cite{Volponi24b}. Adopting graphical user interfaces (GUIs) enables new users to tune control parameters without requiring an in-depth understanding of the underlying code\cite{Keshet13a}. Furthermore, by embedding such a modular system within an unified control architecture that continuously monitors experimental conditions\cite{Barrett22a} and dynamically adapts to system changes\cite{Starkey13a}, full automation of complex experiments can be achieved. This degree of automation not only enhances operational efficiency but also enables experiments to run autonomously outside working hours without the need for human intervention, significantly improving their duty cycle.\\
Various scientific communities have devised control systems and automation protocols tailored to the needs of their experiments. For example, the above mentioned design strategies have been implemented in ultracold atoms \cite{Keshet13a, Barrett22a, Laustsen21a, Vendeiro22a} and accelerator-based experiments\cite{Volponi24a, Volponi24b, Shalloo20a, Yu19a}, automated identification of operating conditions for semiconductor qubits have been reported for quantum-dots setups\cite{Schuff24a, Baart16a, Darulova20a, Durrer20a, Moon20a} and robust number-resolved loading of ion Coulomb crystals have been realized in ion-trapping systems \cite{Kamsap17a, Zawierucha24a, Schmid22a}.\\
Precision spectroscopy of trapped molecular ions is yet another field that can greatly benefit from automation. With the aim to determine fundamental constants with unprecedented accuracy\cite{Patra20a, Kortunov21a}, to test theories on physics beyond the standard model\cite{Safronova08a, Germann21}, and to develop mid-infrared frequency standards unattainable with atomic species\cite{Acef97a}, the level of control over molecular ions and the attainable measurement precision have been steadily improved. While currently a record measurement precision of order $10^{-12}$ has been achieved for such systems \cite{Patra20a, Kortunov21a, Alighanbari25a}, ongoing efforts aim to extend the precision limit further to the one attained for neutral molecules \cite{Shelkovnikov08a} and beyond.\\
Many experimental setups for molecular precision spectroscopy require the repeated preparation of the molecular sample, either because destructive measurement methods\cite{Germann14,Patra20a, Kortunov21a} are employed or because the chemical lifetime limits the effective measurement times \cite{Sinhal20a}. Automated workflows in sample preparation can boost experimental efficiency and ensure consistency while enabling human experimenters to focus on strategic tasks. Another opportunity for automation lies in the execution of frequency scans required for initial identification of spectral lines. For most molecules other than hydrogen isotopologues, theoretical predictions of spectroscopic line positions are significantly less precise than the experimentally achievable precision\cite{Germann14}. This requires extensive scans across broad frequency ranges to initially locate spectral features. Imperfect molecular state preparation and the presence of different nuclear spin isomers can further complicate this process. An automated control system capable of scanning wide frequency ranges, performing real-time data analysis, and adaptively responding to detected spectral features is essential to overcome these obstacles. In addition, automated troubleshooting routines such as laser relocking or compensation of drifting experimental conditions are crucial to maintain long-term stability and prevent measurement downtime.\\
In this article, we present a fully automated control system designed to address common challenges in quantum-logic-spectroscopy experiments with molecular ions. Section \ref{sec:setup} describes our experimental setup for rovibrational spectroscopy of trapped N$_2^+$ molecules. Section \ref{sec:control} introduces the architecture of the control framework that enables fully automated operation of the experiment. Section \ref{sec:routines} details how individual autonomous workflows such as ion loading, dark-ion recognition, ion reduction, molecular mass and state detection, as well as frequency scans, have been implemented. Finally, section \ref{sec:performance} evaluates the robustness and the impact of the automation on the duty cycle of the experiment.

\begin{figure*}
\includegraphics[width=1\textwidth]{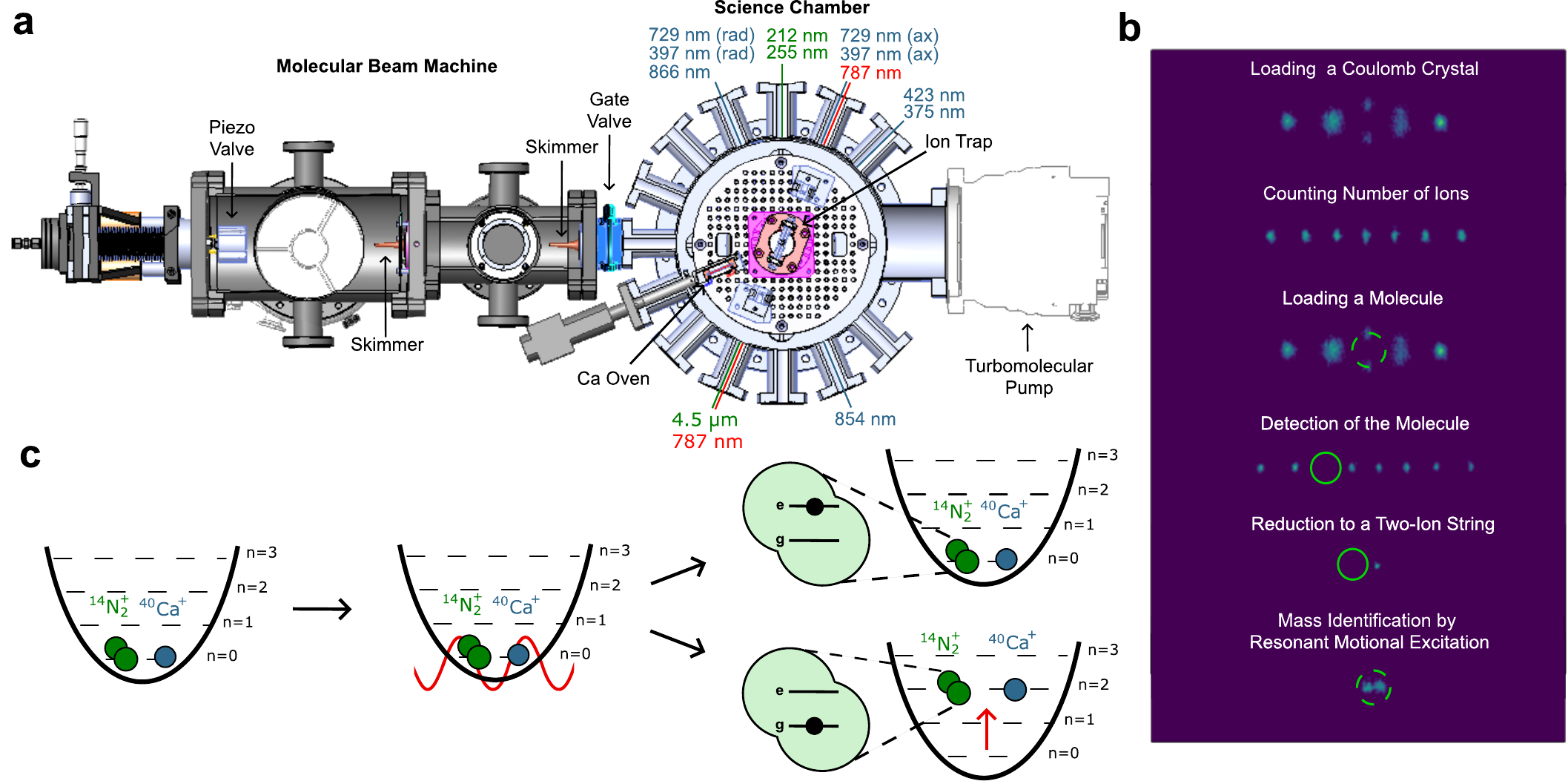}
\caption{\label{fig:setup}a) Experimental setup consisting of a molecular-beam machine and an ultrahigh-vacuum chamber containing the ion trap and a Ca oven. A piezo valve produces a beam of neutral N$_2$ molecules which passes through skimmers and a gate valve to reach the ion trap. Laser beams: blue: photoionization of Ca (423~nm, 375~nm), coherent manipulation and  cooling of Ca$^+$ (729~nm, 397~nm, 866~nm, 854~nm); green: photoionization of N$_2$ (212~nm, 255~nm), spectroscopy of N$_2^+$ (4.5~$\mu$m); red: state detection of N$_2^+$. b) Generation of a Ca$^+$-N$_2^+$ two-ion string: A small Coulomb crystal of Ca$^+$ ions is prepared in the trap. After generation and sympathetic cooling of a single N$_2^+$ ion, the 3D crystal is transformed into a string to more easily detect the single dark molecular ion. Following reduction of the crystal to a two-ion string, the mass of the molecular ion is measured by resonant motional excitation. c) Quantum-logic spectroscopy scheme: a travelling optical lattice exerts an optical dipole force on N$_2^+$ to cause motional excitation of the ion string depending on its internal state. See text for details.}
\end{figure*}

\section{Experimental setup} \label{sec:setup}

The experimental apparatus combines a molecular-beam machine with a linear radiofrequency (RF) ion trap as depicted in figure \ref{fig:setup}a \cite{Meir19a}.  The molecular-beam machine is composed of a pulsed piezo-actuated gas valve (MassSpecpecD ACPV2) and two vacuum chambers at operation pressures of $4\times10^{-6}$~mbar and $2\times10^{-8}$~mbar connected by a skimmer. Another skimmer and a pneumatic gate valve connect the molecular beam machine to an ultrahigh-vacuum science chamber (operation pressure $1\times10^{-11}$ mbar). The valve generates a pulsed supersonic molecular beam of vibrationally and rotationally cold N$_2$ molecules at a repetition rate of 10 Hz.\\
N$_2$ molecules from the beam are ionized at the center of the science chamber using the output of two Nd:YAG-pumped pulsed dye lasers (NarrowScan - Radiant Dyes, Pulsare - FineAdjustment) at 255~nm and 212~nm by a [2+1'] resonance-enhanced multi-photon ionization (REMPI) scheme \cite{Shlykov23a}. The ionization is synchronized with the arrival of the pulsed molecular beam at the center of the ion trap, and laser frequencies and intensities are tuned to produce a single N$_2^+$ ion in its rovibronic ground state. The ion trap provides axial and radial trapping frequencies of approximately 625 kHz (675 kHz) and 755 kHz (1050 kHz) for Ca$^+$ (Ca$^+$-N$_2^+$, in-phase motional mode), respectively. Auxiliary electrodes allow fine control of the trapping potentials for micromotion compensation. Details of the trap can be found in reference \onlinecite{Meir19a}.\\
Before introducing N$_2^+$, a Coulomb crystal of laser-cooled Ca$^+$ ions is prepared. Neutral calcium atoms from a resistively heated oven (AlfaVakuo) inside the vacuum chamber are ionized in the trap center using laser beams at 423~nm and 375~nm and laser-cooled using laser beams at 397~nm and 866~nm. The laser-cooled Ca$^+$ ions sympathetically cool the co-trapped single N$_2^+$ ions. The number of Ca$^+$ ions is then reduced by a controlled lowering of the trap potential to obtain a Ca$^+-$N$_2^+$ two-ion string \cite{Meir19a}. The mass of the molecular ion is then confirmed by mass-dependent resonant motional excitation of the two-ion string using an oscillating electric field applied to one of the trap electrodes \cite{Drewsen04a}. The loading sequence of the dual-species string is depicted in figure \ref{fig:setup}b.\\
Resolved-sideband cooling of Ca$^+$ using laser beams at 729~nm and 854~nm prepares the ion string close to the ground state of its in-phase motional mode\cite{Meir19a}. The state of the molecule is determined non-destructively using a quantum-logic-spectroscopy (QLS) scheme \cite{Sinhal20a}. For this purpose, a travelling optical lattice composed of two counterpropagating laser beams at 787.450~nm modulated by the axial trap frequency applies a state-dependent optical dipole force (ODF) on the ions. At this wavelength, the ODF causes a resonant motional excitation of the ion string if N$_2^+$ is in its rovibrational ground state (see reference \onlinecite{Sinhal20a} for details). By performing Rabi sideband thermometry on the Ca$^+$ ion, information on the motional state of the ion string and hence the molecular state can be extracted (figure \ref{fig:setup}c). Spectroscopy experiments can be implemented with the same QLS scheme by applying additional laser beams on the ion and performing subsequent state-detection measurements to verify successful spectroscopic excitations.\\

\begin{figure*}
\includegraphics[width=\textwidth]{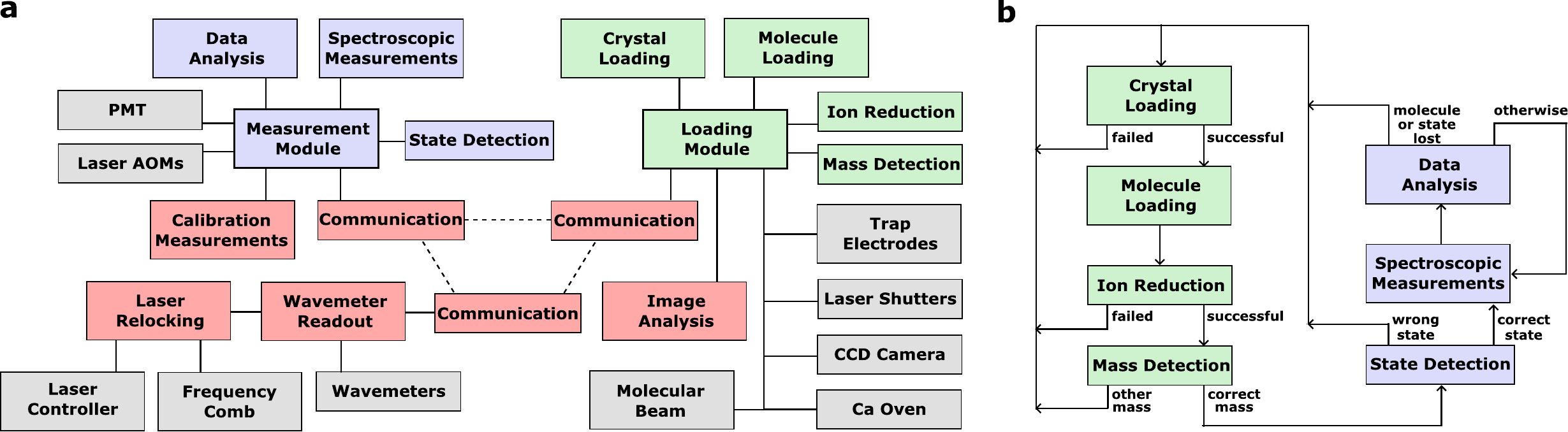}
\caption{\label{fig:control_system} a) The control system comprises an ion-loading module (green) and a measurement module (blue), each controlling a set of hardware components (gray) via software routines for designated tasks. Subroutines (red) such as wavemeter readout and laser relocking support the routines. Communication between different components of the control system (dotted lines) is carried out via TCP/IP. b) The control logic executes routines sequentially, only proceeding if a routine terminated successfully and restarting the cycle otherwise.}
\end{figure*}

\section{The control system} \label{sec:control}

The control system is depicted schematically in figure~\ref{fig:control_system}a. It is organized around two top-level modules: one dedicated to sample loading (green panels in figure \ref{fig:control_system}a) and the other dedicated to conducting measurements (blue panels in figure \ref{fig:control_system}a). Each module controls multiple hardware devices (gray panels in figure \ref{fig:control_system}a) and contains software routines designed to accomplish a specific experimental goal, for instance loading a Coulomb crystal, verifying the mass of the molecule, or performing a state-detection measurement. These routines, in turn, can invoke subroutines (red panels in figure \ref{fig:control_system}a) responsible for simpler or more specialized tasks. During autonomous execution, only one module has full supervisory control at any given time. For instance, the sample loading module will coordinate routines until it declares its objective of loading a Ca$^+-$N$_2^+$ string as completed. Control then passes to the measurement module, which orchestrates execution and analysis of measurements before handing supervisory authority back to the loading module when reloading of ions becomes necessary again. This modular design with well-defined routines and subroutines combined with GitHub-based version control of all involved scripts allows for upgrades of algorithms and integration of additional devices with minimal disruption to the running experiment. In addition, each routine is linked to a graphical user interface, which enables adjustments of control parameters in real time and without knowledge of the underlying code. This is particularly useful for calibration and testing of new features and for multi-user experiments. Since each routine can be executed and controlled either by a user via a GUI or autonomously by one of the two top-level modules, fully autonomous, semi-autonomous, or completely manual modes of operating the experiments are available.\\ 
The measurement module contains routines executing experimental sequences that require microsecond-precise synchronization of multiple acousto-optical modulators (AOMs) with a photomultiplier tube. 
Such time-critical experiments are implemented on the Sinara/ARTIQ ecosystem, which combines open-source control electronics with a Python-based programming language to handle the communication between a host computer and a field-programmable gate array (FPGA) with nanosecond timing resolution and sub-microsecond latency\cite{Artiq}. This platform enables the realization of experimental sequences composed of dozens of precise laser pulses necessary, for instance, for the ground-state cooling of ion strings or Rabi thermometry \cite{Sinhal20a}. All other routines, for instance, for data analysis or the routines from the sample loading module, operate on timescales ranging from milliseconds to minutes and thus do not require specialized electronics. These routines are either implemented with LabVIEW\cite{Labview} or with Python\cite{Python} for the orchestration of the overall automation logic and the execution of intricate analysis algorithms.\\
Figure \ref{fig:control_system}b shows a flow chart of the main control logic. The supervisory scripts for the loading and measurement modules advance sequentially through individual routines. In the loading module, the sequence begins an initialization procedure (see section~\ref{sec:loading}) and then continues with the generation of a Coulomb crystal of Ca$^+$ ions, trapping and sympathetically cooling a single N$_2^+$ ion, reducing the crystal to a two-ion chain, and verifying the mass of the molecular ion. The success of each step is determined through on-the-fly image analysis of a CCD camera feed (see figure \ref{fig:setup}b for example images). If one of these routines indicates a futile execution, for instance, due to drifting parameters, the system invokes a new loading cycle and attempts to correct the underlying cause by adjusting the relevant control parameters. Only after the loading module declares successful loading of a Ca$^+-$N$_2^+$ string does the measurement module assume supervisory control.\\
The measurement module starts by determining the internal state of the molecule using QLS. If the molecule is detected in its ground state, spectroscopic measurements in the form of excitation pulses with subsequent state-detection measurements are performed. After every measurement, the data obtained is analyzed and interpreted allowing the system to autonomously react in order to repeat the experiment with more precise scans around the corresponding frequencies if needed. Eventually, the molecule itself or its state will be lost by collisions or off-resonant photon scattering which triggers the loading module to take over again and restart the ion-loading protocol.\\
As the experiment spans across multiple optical tables as well as multiple computers and incorporates dozens of distinct hardware devices, a distributed layout of the control system is inevitable. Communication between the various programs and devices occurs through message-passing over TCP/IP connections structured similarly as proposed in the actor model\cite{Actor_model}. Whenever a particular routine or subroutine is instructed to perform a task, a command is sent to the relevant process. The message is then echoed back to the sender to prevent lost or corrupted instructions. After accomplishing its task, the actor sends back a confirmation message and awaits new instructions. Many routines and subroutines are additionally monitored by watchdog timers, which are parallel processes that stop the experiment if a routine or subroutine stalls or critically fails. 
With watchdog timers implemented, robust communication between different scripts, and self-correcting mechanisms built into the top-level control algorithms, uncontrolled system failure is designed to be prevented. If one of these three mechanisms signals a critical failure, for instance, due to multiple consecutively failed attempts for the same routine or subroutine, or, a timeout event due to some unforeseen occurrences, the entire experiment is suspended in a controlled and safe manner.\\
Throughout the whole experimental cycle, relevant diagnostic information is recorded in real time. For instance in the loading module, the outcome of each routine with the values of the relevant control parameters and optionally the acquired camera images are logged with timestamps. ARTIQ-based measurements, on the other hand, are assigned a unique execution ID and the measurement results are recorded alongside the experimental configuration. 
Timestamped logs of the laser frequencies, of relocking or recalibration events, and of the real-time analysis outcomes allow the correlation of any spectroscopic anomalies with the corresponding laser properties. This thorough logging strategy enables full reconstruction of any autonomously made decision and facilitates posterior data analysis. Thus, unsupervised automated sample loading, molecular state detection and spectroscopic measurements are achieved over long time periods with the presented control system.

\section{Automated Routines} \label{sec:routines}
\subsection{Ca$^+$ Crystal Loading and Analysis} \label{sec:loading}
Upon invoking the crystal loading routine, an initialization subroutine is called first. This initialization begins by releasing any species remaining in the ion trap from previous experiments. Heating of the Ca oven is then initiated, which usually takes about one minute and represents the longest task in the experimental protocol. In the meantime, another subroutine verifies whether the 729~nm laser is properly locked to an ultrahigh-finesse cavity and automatically relocks it if necessary. Next, all involved wavemeters are calibrated against the 729~nm master laser to ensure there are no frequency offsets in subsequent wavemeter readings. The frequencies of all other lasers are then checked and, if required, relocked (see section \ref{sec:lasers}).\\
Once the oven reaches a temperature that yields an adequate calcium flux, shutters blocking the oven and the lasers for ionization and for laser cooling of the Ca$^+$ ions are opened. Crystal loading proceeds by iteratively verifying the size and position of the Coulomb crystal via a real-time image analysis subroutine illustrated in figure~\ref{fig:image_analysis}. Noise in the raw CCD images is first reduced using a Gaussian smoothing filter, and intensity peaks along horizontal slices through the images corresponding to single ions are then identified using a dynamic threshold derived from the background level and their topographic prominence (the height of a peak relative to the surrounding minima) to distinguish ions even in blurred crystal images (see figure~\ref{fig:image_analysis}). Guided by the position and size of a reference crystal from a prior calibration, the horizontal slice exhibiting the largest crystal extent of the current image frame is located. The distance between the outermost ions on this slice, averaged over several frames to account for fluctuations, defines the horizontal crystal size. If the crystal is smaller than a predefined target size, oven heating and photo-ionization continue. If the crystal loading procedure overshoots the target size, the RF potential is temporarily lowered multiple times to remove surplus ions.\\
On rare occasions, no bright ions appear within a given time after starting the loading routine. Extended autonomous operation can accumulate patch potentials if the ionization lasers are not well aligned, shifting ion positions and increasing micromotion. In such cases, heating rates can outpace cooling rates, particularly when multiple ions enter the trap rapidly. Additional RF modulation during loading has proven effective at removing hot ions and helping to capture ions that would otherwise remain outside the stable trapping region. Another challenge is crystals continuing to grow for tens of seconds even after the oven and ionization lasers have been turned off. An immediate temporally limited RF-amplitude modulation once the crystal reaches its target size, coupled with a short waiting time before proceeding, mitigates this overgrowth. Finally, to ensure comparably bright and sharp fluorescence peaks for differently sized ion ensembles, the frequency of the 397 nm cooling lasers and the power ratio between its different beams along the radial and axial directions of the trap (see figure \ref{fig:setup}a) are adjusted automatically according to the crystal size.

\subsection{N$_2^+$ Loading and Dark-Ion Recognition}
\begin{figure}
\includegraphics[width=0.49\textwidth]{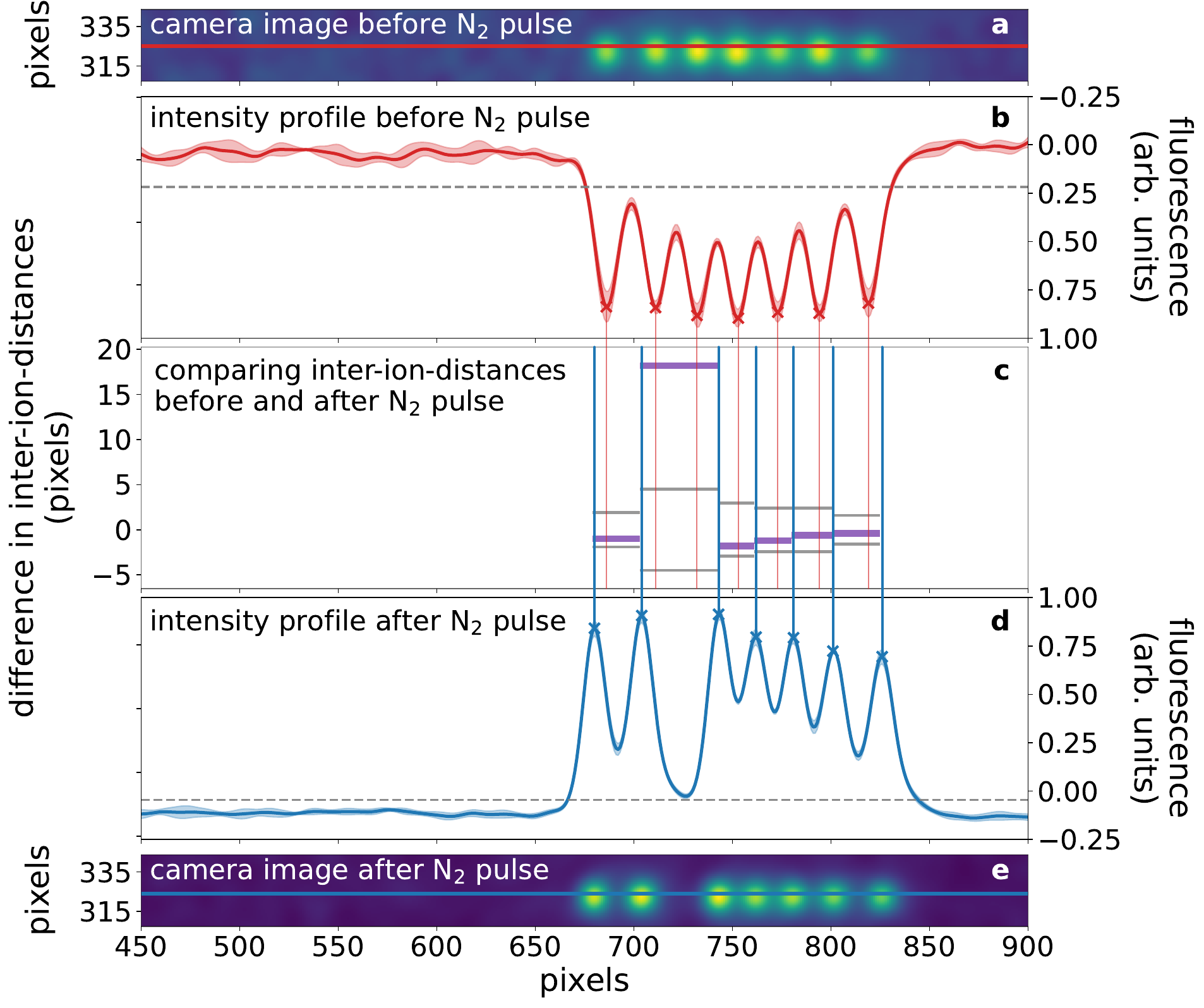}
\caption{\label{fig:image_analysis} Camera images (panels a, e) and corresponding fluorescence-intensity profiles (panels b, d) before and after loading a single N$_2^+$ ion. Shaded areas in (b) and (d) indicate the standard deviation over an average of multiple frames. Vertical red and blue lines designate peak markers, dashed gray lines the peak-finding thresholds. The detector panel (c) visualizes the dark-ion detection algorithm. Purple lines indicate the difference in the inter-ion distances and gray vertical lines mark thresholds $\pm T_i$ for each gap. See text for details.}
\end{figure}

A single N$_2^+$ ion is introduced into the Coulomb crystal by selectively ionizing a neutral nitrogen molecule from the molecular beam using the pulsed dye-laser beams and then relying on the Ca$^+$ ions to sympathetically cool the molecular ion to integrate into the crystal. Because N$_2^+$ is not laser cooled, the molecular ion appears as a dark "defect" in the fluorescence images of the Ca$^+$ ion crystal (figure~\ref{fig:setup}b). Simple autonomous loading approaches such as applying a fixed number of ionization pulses proved unreliable owing to the stochastic nature of each ionization event and the gradual bleaching of dye solutions, which lead to drifting and fluctuating laser-beam intensities. This regularly resulted in loading either no or multiple molecular ions, underscoring the need for a robust dark-ion recognition algorithm.\\
A comparably simple protocol based on generating ion strings was found to lead to a reliable identification of single N$_2^+$ ions embedded in an ensemble of Ca$^+$ ions (figure~\ref{fig:image_analysis}). In this method, the positions and pairwise distances of the bright Ca$^+$ ions within the string are first extracted from CCD camera images averaged over several frames before the first molecular beam pulse. The mean value of the inter-ion distances $\mu_i$ and their standard deviations $\sigma_i$ over these reference frames are determined for each gap $i$ between the fluorescence maxima of the ions along the string. Before each molecular-beam pulse, the axial trap frequency is briefly increased to obtain a three-dimensional ion structure to enhance the sympathetic-cooling efficiency compared to the one-dimensional string. To determine whether a dark ion was loaded, the trap settings are switched back to the string configuration after each pulse and new frames are recorded. In each of these frames, the inter-ion distances $d_i$ are compared individually to the corresponding reference values using the criterion $\lvert d_i - \mu_i\rvert > T_i$, where $T_i=\max(6\sigma_i,1.8)$ is a threshold value for the internal gaps and $T_i=\max(4\sigma_i,1.8)$ for the two gaps of the outermost ions at both ends of the string. The latter positions are tested more restrictively because a dark ion located at one of the ends of the string only has one bright neighbor and shifts the positions of the remaining ions, making its signature more subtle. The lower, empirical bound of 1.8 prevents unrealistically small thresholds if $\sigma_i$ are very small. A dark ion is flagged if the condition is satisfied for a given gap in a majority of the frames, making the algorithm robust to ion hopping or intermittent blurring of the images. In this way, the algorithm yields a reliable classification of dark-ion presence without requiring prior knowledge of its mass. To accommodate optimal cooling in both 1D string and 3D crystal configurations, the 397~nm cooling laser frequency and the power ratio of its axial and radial beams (figure~\ref{fig:setup}a) are adjusted synchronously with the electrode voltages.\\
Table \ref{tab:N2_loading} illustrates typical results from this protocol drawn from a representative measurement campaign. For this purpose, the analysis is restricted to experiments yielding either a Ca$^+$-N$_2^+$ string or a single Ca$^+$ ion. True positives (TP) and true negatives (TN) are repetitions in which the recognition algorithm correctly identified the presence or absence of a dark ion. False positives (FP) are defined as experiments in which the algorithm predicted a dark ion but only a single Ca$^+$ ion was present during mass identification, while false negatives (FN) are measurements in which the algorithm did not predict any dark ion but an N$_2^+$ ion was found to be present nonetheless. Precision (TP/(TP+FP)) quantifies the reliability of dark-ion predictions, while recall (TP/(TP+FN)) measures the fraction of all experiments containing a dark ion that the algorithm successfully identified.\\
Table \ref{tab:N2_loading} also compares the string-based strategy with attempts of applying the same dark-ion recognition algorithm to 3D Ca$^+$ crystals composed of hundreds of ions. The algorithm based on comparing bright-ion positions and distances to a reference proved unreliable for these samples since collisions with the molecular beam frequently induced structural phase changes in the crystals, or trapped N$_2^+$ ions could localize on a crystal plane not imaged by the camera. Despite the slower average sympathetic-cooling time, the string-based protocol achieves significantly higher recall and precision and was thus considered more efficient for fully autonomous N$_2^+$ loading.

\begin{table}
\caption{\label{tab:N2_loading}Dark-ion recognition performance for crystal- and string-based loading strategies.}
\begin{tabular}{|c||c|c|}
\hline
& crystal strategy & string strategy\\
\hline\hline
total runs      & 127 & 2496\\
true positives (TP)& 18 & 1939\\
true negatives (TN)& 54 & 230\\
false positives (FP)& 20 & 270\\
false negatives (FN) & 35 & 57\\
\hline
precision       & 47.4\,\% & 87.8\,\%\\
recall          & 34.0\,\% & 97.1\,\%\\
symp.\ cooling time & $\sim$ 15 s & $\sim$ 30 s\\
\hline
\end{tabular}
\end{table}

\subsection{Ion Reduction} \label{sec:reduction}

\begin{figure}[b]
\includegraphics[width=0.49\textwidth]{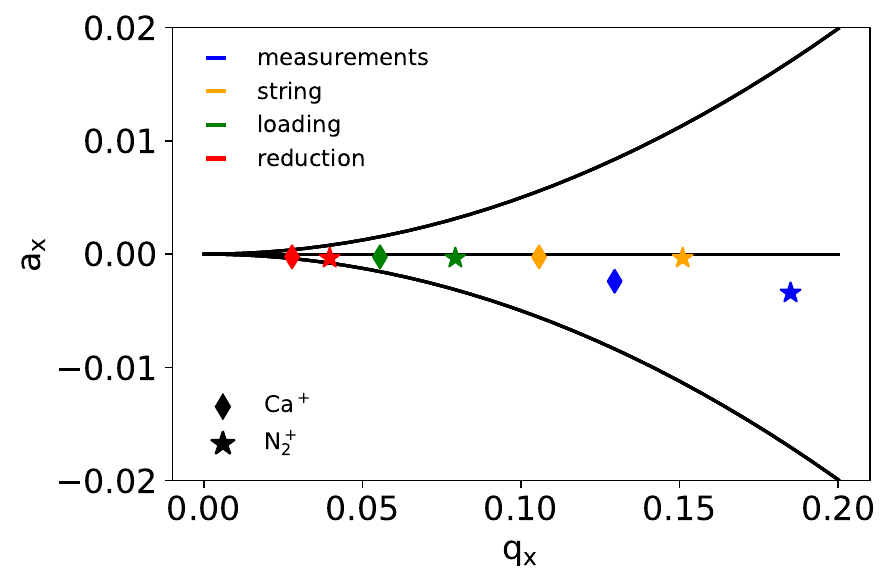}
\caption{\label{fig:stability} Stability diagram of the ion trap indicating configurations used during measurements (blue), ion loading (green), ion counting (yellow), and crystal reduction (red) for Ca$^+$ (diamonds) and N$_2^+$ (stars).}
\end{figure}

To reduce the crystal to a two-ion Ca$^+$-N$_2^+$ string, the control software repeatedly lowers the RF amplitude of the trap for short amounts of time (80 - 160 ms) in order to allow calcium ions to escape the trap. Figure \ref{fig:stability} illustrates the stability diagram in terms of the Mathieu $a$ and $q$ parameters \cite{major05a} of the ion trap for Ca$^+$ and N$_2^+$ at the settings used during ion loading, mass and state detection, as well as the reduction events. Since $^{40}$Ca$^+$ has a heavier mass than $^{14}$N$_2^+$, it experiences a shallower pseudopotential of the trap. Lowering the RF amplitude and thereby decreasing the $q$ parameter pushes Ca$^+$ toward the edge of the stability region while keeping N$_2^+$ well confined. After each RF lowering event, a short settling interval ensures that any lost ions have left the trapping region and that the remaining crystal has stabilized. The software then re-examines the crystal’s size and adjusts control parameters if necessary. In cases where no or only a small percentage of the trapped ions are ejected, the algorithm applies subsequent RF "kicks" at lower amplitudes. Conversely, if too many ions are lost, the setpoint for the RF amplitude kicks is raised to avoid losing the entire crystal within the next reduction step. \\
When only a few bright ions remain, they may appear overlapped as a single broad peak in the camera image, depending on the exact trap configuration. To avoid misidentifying this overlap as a single bright calcium ion, the image analysis routine begins fitting Lorentzians to any detected peaks once only a couple of ions remain. The reduction process concludes only if a single peak is detected whose width matches that of a single trapped ion, after which the trap voltages are adjusted to a configuration optimized for mass and state detection.

\subsection{Mass Identification} \label{sec:tickling}
Up to this point, the chemical nature of the dark ion is unknown. While N$_2^+$ is expected, occasional ionization and trapping of background gas molecules or reaction of N$_2^+$ with the background gas can lead to different ion species being trapped. Mass identification of the dark ion is accomplished by probing the resonance frequency for axial normal modes of the two-ion string. For two ions of masses $m_{\rm Ca}$ and $m_{\rm X}$, the mode frequencies are
\begin{equation}
\omega_{\pm}=\omega_z\sqrt{1+\mu \pm \sqrt{1+\mu^2-\mu}},
\end{equation}
where $\mu=m_{\rm Ca}/m_{\rm X}$ and $\omega_z$ is the single-ion axial secular angular frequency of $^{40}$Ca$^+$ \cite{Morigi01a}.\\
Before measuring the resonance frequency of the two-ion string, the image-analysis subroutine measures the width and brightness of the fluorescence spot of the co-trapped Ca$^+$ ion. This is used as a reference to detect broadening of the spot under the action of resonant motional excitation. The software then iterates through a predefined list of candidate masses, applying the corresponding drive frequency, analyzing the ion images, and allowing a brief stabilization period before moving on to the next frequency.\\
Depending on the driving amplitude and the laser parameters, the bright ion may disappear entirely at the applied resonance frequency. In such instances, the code attributes the disappearance to resonance effects rather than a genuine ion loss. To verify the resonance, the drive is temporarily disabled and it is checked whether the bright ion reappears immediately displaying its original fluorescence profile. If no resonance is observed within the list of expected masses, the code optionally checks for additional possible values of the mass. In most cases, however, the absence of a resonance within the expected masses is explained by the presence of multiple dark ions, in which case the trap is emptied and the loading procedure reinitialized.\\

\subsection{Molecular-State Detection}
\begin{figure}
\includegraphics[width=0.49\textwidth]{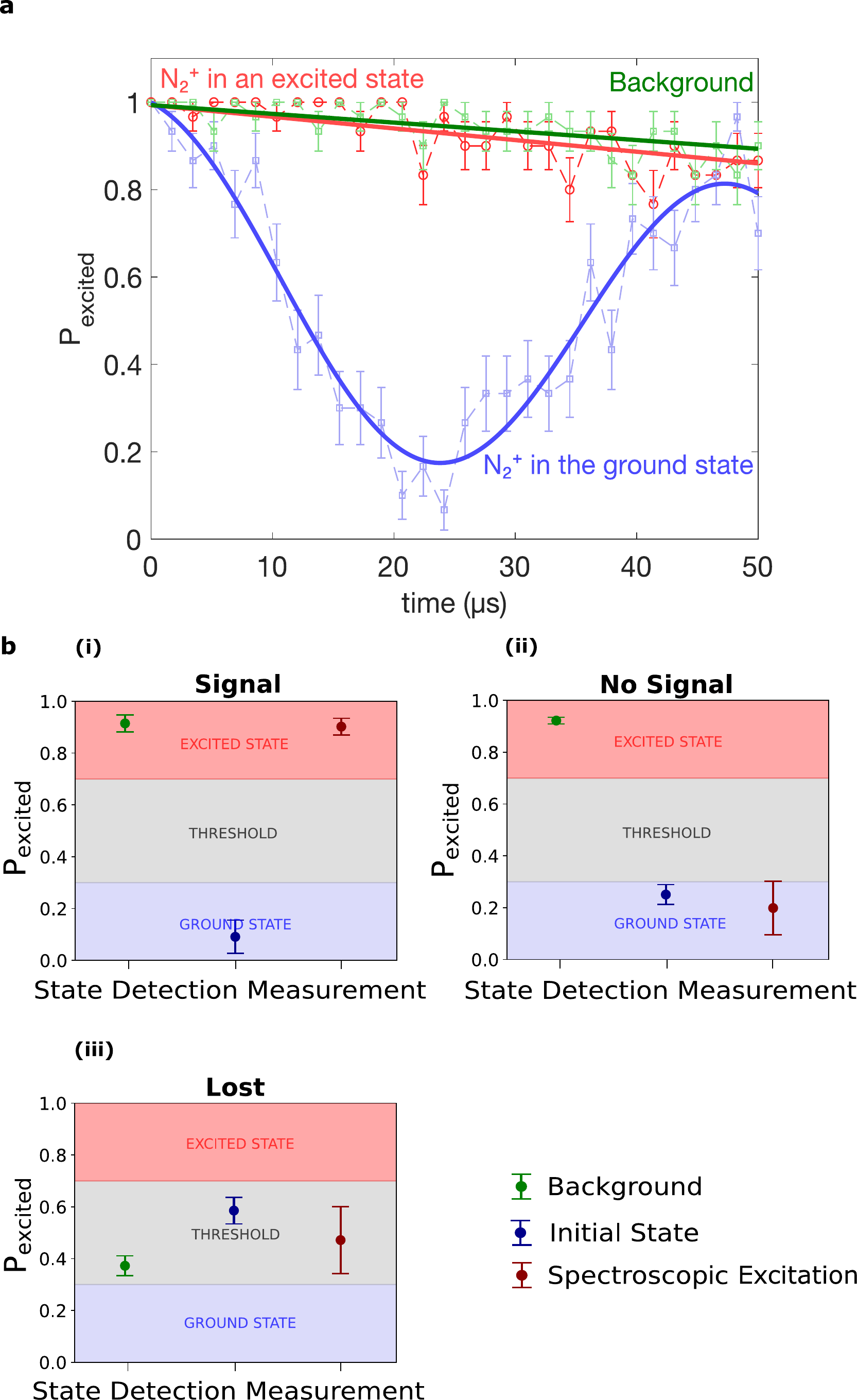}
\caption{\label{fig:state_detection} a) Rabi flop on a motional sideband of the in-phase motional mode for N$_2^+$ in its rovibrational ground state (blue trace) and in an excited state (red trace). A background measurement before applying the lattice lasers  (green trace) serves to discriminate systematic errors. Solid lines represent fits of the data \cite{Sinhal20a}. b) In subsequent QLS measurements, only data points around the time of maximum contrast in the Rabi flop are taken, averaged to a single value and compared with each other yielding the following outcomes of the state-detection measurements during the spectroscopy protocol: (i) successful spectroscopic excitation experiment from the ground state, (ii) no spectroscopic excitation detected, (iii) loss of ion during the sequence. See text for details.
}
\end{figure}

While the routines within the loading module run on comparatively slow timescales (seconds to minutes), state detection via QLS \cite{Sinhal20a} relies on microsecond-precise control of laser pulses implemented within the ARTIQ framework. First, resolved-sideband cooling of the axial in-phase mode of the Ca$^+-$N$_2^+$ string close to its motional ground state is performed on the 729~nm clock transition in Ca$^+$. The Ca$^+$ ion is then optically pumped into its 3d$^2\text{D}_{5/2},\, m = -5/2$ Zeeman state to suppress motional excitation caused by spurious ODF on the Ca$^+$ ion during the subsequent exposure to the lattice lasers \cite{Sinhal20a, Najafian20b}. Following state-dependent motional excitation of the ion string by the ODF, measurements of Rabi oscillations on the blue sideband of the in-phase motional mode associated with the 729 nm transition in Ca$^+$ distinguish between a N$_2^+$ molecule in its rovibrational ground state versus an excited state as shown in figure \ref{fig:state_detection}a \cite{Sinhal20a}. A background measurement of the same Rabi oscillation without the lattice lasers, but with a wait time equal to the duration of the coherent motional excitation, is conducted to characterize the efficiency of motional-ground-state cooling and decoherence effects like anomalous heating. If N$_2^+$ is in its internal ground state, the amplitude and frequency of the Rabi oscillation are extracted from suitable fits \cite{Sinhal20a}. This allows to shorten the time spent during future state detection measurements by only recording data points around the Rabi time where the flops exhibit the highest contrast between N$_2^+$ in its rovibrational ground state and in an excited state.\\
To ensure a sufficiently high contrast over long periods in the presence of potentially drifting parameters, calibration measurements are performed about every hour of autonomous operation. Since the ground-state cooling efficiency of the Ca$^+-$N$_2^+$ string does not depend on the internal state of the molecule and every N$_2^+$ ion in its rovibrational ground state should be used for spectroscopic experiments, these calibration measurements are performed whenever an N$_2^+$ ion was found to be spuriously generated in an excited state \cite{Shlykov23a} and enough time has passed since the last calibration measurement. These measurements include spectra and Rabi-frequency measurements of Zeeman transitions within the manifold of the 729 nm clock transition in Ca$^+$ to ensure efficient ground-state cooling and optimal pumping to the $\text{D}_{5/2},\, m = -5/2$ state.

\subsection{Molecular spectroscopy} \label{sec:spectro}
A key application of the present experiment is molecular spectroscopy. Once N$_2^+$ has been prepared in its rovibrational ground state and addressed by the preceding protocols, the ARTIQ framework manages subsequent spectroscopic measurements. \\
Here, the primary objective of the automated measurement sequences is to identify the exact frequency of the transitions which necessitates a real-time analysis of the experiments to accurately determine the presence of a genuine spectroscopic excitation signal. The spectroscopy procedure thus follows a systematic approach in which prior to any spectroscopic excitation, background and state-detection measurements are performed to ensure that the molecular ion is prepared its rotational ground state (figure \ref{fig:state_detection}a). Next, a pulse from the spectroscopy laser, tuned to a specific frequency, is applied to attempt excitation from the ground to a specific excited state. This is immediately followed by another state-detection measurement to check if the interaction has altered the internal state of the molecular ion. Subsequently, a second spectroscopy pulse is applied at the same frequency, again followed by state detection to ascertain whether the ion returned to the ground state and is thus re-initialized. The algorithm ensures that N$_2^+$ is prepared in its ground state again by applying, if necessary, several de-excitation pulses before proceeding with the frequency scan. The outcome of a successful spectroscopic excitation event is shown in figure~\ref{fig:state_detection}b (i). If no transition is detected (panel (ii) in figure \ref{fig:state_detection}b), the sequence is repeated with a different laser frequency until an excitation is observed.\\
The second objective of automating the spectroscopy procedure is to enhance both speed and reliability. In the present experimental environment, N$_2^+$ ions have a finite chemical lifetime due to reactions with residual H$_2$ molecules in the vacuum. With a background pressure of order $10^{-11}$ mbar, the typical chemical lifetime of N$_2^+$ is on the order of tens of minutes limiting the available experiment time before the molecular ion is lost. A typical signature of a measurement in which N$_2^+$ has been lost during the experimental sequence is shown in panel (iii) of  figure~\ref{fig:state_detection}b. \\
Additionally, the laser used for state detection can cause off-resonant scattering that drives unwanted electronic transitions in the molecular ion, after which the initial state of the molecule can usually not be recovered. By optimizing the sequence for speed, a broader frequency range can be scanned before N$_2^+$ is either chemically destroyed or its internal state is irretrievably lost. Finally, different sets parameters for the spectroscopy sequence can be loaded and automatically managed by ARTIQ, which logs the results in the appropriate files. This automation significantly reduces human errors that can cause data inconsistencies.

\subsection{\label{sec:lasers}Laser Stabilization}
A total of 11 different laser sources are used in the present experiment, each stabilized with varying levels of precision depending on its purpose. Three different types of locking schemes are employed to ensure the required frequency stability.\\
The most critical laser, serving as the master reference for the system, is the external-cavity diode laser addressing the Ca$^+$ clock transition at 729 nm (see section \ref{sec:setup}). This laser is locked  to an ultra-low-expansion (ULE) cavity (Stable Lasers) with a quoted finesse of 270,000 using the Pound-Drever-Hall (PDH) \cite{Pound–Drever–Hall} technique. The lasers used for ionizing calcium (423~nm) and N$_2^+$ (255~nm), repumping Ca$^+$ population (866~nm and 854~nm) and enabling molecular-state detection (787~nm) are connected to wavemeters (HighFinesse WSU-30 and WS6) which are regularly calibrated using the 729~nm master laser. \\ 
A second category requiring more precise locking includes the lasers for Doppler cooling and imaging of Ca$^+$ (397 nm) and for N$_2^+$ spectroscopy. These lasers are stabilized by a lock to an optical frequency comb (OFC, MENLO FC1500-250-ULN). This is implemented by generating beat notes of the laser and OFC outputs which are then processed by Mixing Fast Analog Linewidth Control (mFALC) modules (Toptica). Since these lasers can occasionally lose their lock, an automated relocking scheme is implemented. The relocking routine iteratively adjusts the piezoelectric actuator to steer the laser frequency toward the target beat frequency and subsequently verifies the detuning between the laser and the corresponding comb tooth by modulating the local oscillator. The repetition rate of the OFC is itself locked to the 729~nm master laser for narrowing the linewidth of the individual comb teeth. Slow variations of the repetition rate caused by drifts of the ULE cavity are compensated by a remote reference to the Swiss primary frequency standard via a stabilized telecom fiber link \cite{Husmann21a}. \\ 
The final group of lasers includes those used for ionizing both particles: 375~nm for calcium and 212~nm for N$_2^+$. These lasers are not frequency locked with high precision but are crucial for initiating the ionization process. Only the latter laser (212~nm) is monitored and compensated manually if needed. \\
Environmental factors, such as temperature and humidity fluctuations, can affect laser stability and the accuracy of wavemeter readings. To mitigate these effects, the wavemeters are recalibrated at the start of each experiment during the generation of the Ca$^+$ ion crystal (see section \ref{sec:loading}).\\
Given the critical role of the 729 nm master laser in stabilizing the entire system, its lock to the ULE cavity must be continuously monitored. To ensure it remains locked to the correct cavity mode, the laser voltage is slightly modulated, and the frequency response is analyzed. If the reading shows instability due to the modulation, an automated relocking process is triggered. This involves releasing the PDH lock, managed by the FALC, and re-engaging it at the correct frequency.\\
When the 729~nm laser loses lock, a cascading relocking sequence is initiated to ensure all dependent lasers are correctly re-stabilized. Both the loading and measurement modules frequently check the status of the master laser to detect and respond promptly to any loss of lock, maintaining overall system stability.

\section{Safety measures} \label{sec:safety}
Since the control system is capable of unsupervised autonomous operation for prolonged periods, it must include safeguards for handling unforeseen events beyond compensation of drifting parameters and self-correcting control parameters within individual routines. Several software- and hardware-level measures are therefore implemented to ensure that the experiment aborts in a controlled manner if unexpected failures arise and no damage to critical components happens.\\
On the software side, message integrity is ensured by echoing every TCP/IP command, allowing the supervisory control script to confirm that each routine or subroutine follows the intended instructions. Furthermore, watchdog timers monitor the duration of each routine and intervene whenever a routine or its subroutines run longer than expected as this may indicate a hardware failure or a communication breakdown. If a single routine fails repeatedly despite parameter adjustments, the entire experiment transitions into a controlled shutdown. This shutdown protocol involves closing laser shutters and stopping the Nd:YAG laser, discontinuing the calcium oven and the molecular beam machine, switching off all voltages applied to the ion trap, and terminating any remaining active routines. In addition, operators maintain the highest level of control authority with the ability to manually interrupt the autonomously running experiment at any time.\\
Additional hardware-oriented safeguards include an independent system that continuously tracks the vacuum pressure in the main chamber. Whenever predefined thresholds are exceeded, the system sends e-mails and text notifications to operators alarming them to take action. Because restoring stable ultrahigh-vacuum requires extensive downtime, redundant safety measures for the two devices most likely to impair vacuum conditions (the calcium oven and the pulsed piezo valve) are implemented with additional hardware-based watchdog timers. Lastly, the entire setup is connected to uninterruptible power supplies (UPS) to maintain pump operation and messaging capabilities even during power outages.

\section{Performance} \label{sec:performance}
One of the main objectives of the present automated control system is to reliably produce Ca$^+$–N$_2^+$ strings with N$_2^+$ in its rovibrational ground state. The total success rate for this objective directly depends on the success rate of the constituent routines in the loading module as well as on the efficiency of the REMPI ionization scheme to produce ground‐state molecules. To evaluate the individual routines, a sample of 5578 loading cycles collected over three months was analyzed.\\
The first relevant routine, crystal loading, succeeded 95.2\% [94.6\%, 95.7\%] of times (numbers square brackets indicate the 95\% Wilson confidence interval\cite{Brown01a}), failing only if excessive micromotion hinders efficient laser cooling. Reduction was successful in 85.9\% [84.9\%, 86.8\%] of cases with slow drifts of the RF amplitude being the main cause of total ion loss. In 60.3\% [58.8\%, 61.7\%] of cases, tickling identified one of the three tested masses (28, 29, or 40 amu). In all other cases, either a dark ion of a different mass or two dark ions were present. Out of the 5578 loading cycles, the successful loading of a single N$_2^+$ ion was verified in 1996 cases, leading to an overall success rate of preparing a Ca$^+-$N$_2^+$ string of 35.8\% [34.54\%, 37.1\%].\\
The fidelity of the QLS state‐detection scheme was previously reported to exceed 99\% \cite{Sinhal20a}, and the efficiency of the REMPI scheme to produce ground state molecular ions was reported to be 38$\pm$7\% \cite{Shlykov23a}. Of the 1996 autonomously loaded N$_2^+$ ions in the present study, 26.1\% [24.3\%, 28.1\%] were found in the rovibrational ground state. Deviations from the value reported in reference \onlinecite{Shlykov23a} are attributed to fluctuations in the alignment of the REMPI lasers during autonomous operation which has a critical influence on the state-preparation fidelity \cite{Shlykov23a}. Taking the efficiency of the REMPI scheme into account, the total success rate of autonomously loading a Ca$^+-$N$_2^+$ string with N$_2^+$ in its ground state is 9.3\%.\\
The duration of a single loading cycle is 241 $\pm$ 91 seconds (standard deviation). This time is largely limited by unavoidable hardware constraints such as heating up the Ca oven that can last for up to a minute, and the statistical nature of N$_2^+$ loading involving a 30 seconds waiting time after every loading attempt to ensure proper sympathetic cooling. With these constraints, the runtime of the experiment in autonomous operation is comparable to manual operation. However, autonomous loading of ions still provides advantages over manual loading by freeing staff from repeated tasks, enhancing reproducibility due to the standardized protocols, and extensive data logging which proves useful during troubleshooting. \\
As discussed in section~\ref{sec:spectro}, the finite state and chemical lifetimes of N$_2^+$ restrict the number of measurements obtainable from a single molecule. The autonomous control system reduces idle times between measurements. Moreover, automated parameter control prevents operator mistakes during repetitive scans, which further helps to make efficient use of the limited molecular lifetime. Since both loading and measurement modules can run completely unsupervised, the experiment’s uptime naturally extends. In practical usage monitored over a three-month period, nearly a factor of 10 increase in processed experimental cycles and about a factor of 8 increase in the number of loaded molecules was achieved compared to manual operation over the same time span before the automated control system was deployed.\\
Despite these benefits, occasional human intervention remains necessary. The dye solutions used for the photoionization lasers degrade over time requiring frequent optimizations of laser parameters and, eventually, replenishment. Other effects necessitating manual re-optimization are micromotion compensation, mitigation of slow drifts in laser alignments and occasional triggering of titanium sublimation pumps to maintain good vacuum conditions. In principle, also these tasks could be included in the automation. Typically, the experiment can currently run fully autonomously for about two days before any human intervention is required.\\

\section{Conclusion}
The automated control system presented in this work was motivated by the increasing complexity of modern molecular-ion experiments, where frequent sample preparation and the need for precise timing place substantial demands. Without a robust, automated solution, measurement run times and rapid, reliable reloading of ions would remain major bottlenecks in experiments aiming for high duty cycles and state-of-the-art precision. Our implementation surmounts these challenges by integrating modular workflows for crystal loading, dark ion recognition, ion reduction, and mass as well as state detection of single molecular ions. Through real-time image processing and adaptive feedback loops, the system can maintain unsupervised operation over extended periods of time, increasing the total number of experimental cycles by nearly a factor of ten, while the total number of loaded molecules increased by about a factor of eight compared to manual operation.\\
Future extensions may reduce drifting parameters even further to improve molecular-ion-loading success rates. In particular, more stable lasers for photoionization of N$_2$ would alleviate major fluctuations, and autonomous micromotion compensation performed regularly would maintain optimal trapping conditions over extended uninterrupted measurement campaigns. Advanced algorithms for dark-ion recognition, potentially incorporating machine learning, also hold promise for more efficient ion detection \cite{Yin25a}. 

\section*{Acknowledgements}

The present work was supported by the Swiss National Science Foundation (grant nr. 200021\_204123), the European Partnership on Metrology (Funder ID: 10.13039/100019599, grant nr. 23FUN04 COMOMET), and the University of Basel.

\newpage
\bibliography{references}

\end{document}